%% file: main.tex
\documentclass[sigconf,screen]{acmart}

\usepackage{amsthm}
\usepackage{algorithm}
\usepackage{textcomp}
\usepackage{stfloats}
\usepackage{verbatim}
\usepackage{graphicx}
\usepackage{multicol}
\usepackage{multirow}
\usepackage{siunitx}
\usepackage{comment}
\usepackage{bm}
\usepackage{wrapfig}
\usepackage{enumitem}
\usepackage{cleveref}
\usepackage{rotating}
\usepackage{booktabs}
\usepackage{balance}
\usepackage{fancyhdr}
\usepackage{bbm}
\usepackage{soul}
\usepackage{mathtools}
\usepackage{pifont}
\usepackage{makecell}
\usepackage{subcaption}
\usepackage{xspace}
\usepackage[skins]{tcolorbox}
\usepackage{etoolbox} 

\setlist{noitemsep, leftmargin=*, topsep=0pt, partopsep=0pt}

\crefformat{section}{\S#2#1#3} 
\crefformat{subsection}{\S#2#1#3}
\crefformat{subsubsection}{\S#2#1#3}
\AtBeginDocument{%
  }

\setcopyright{none}
\renewcommand\footnotetextcopyrightpermission[1]{}

\makeatletter
\newenvironment{takeaway} {\par\vspace{-0.1em}\begingroup\refstepcounter{takeaway}\textbf{Takeaway \thetakeaway.}~}
  {\par\endgroup}
\newcounter{takeaway}
\makeatother

\copyrightyear{2025}
\acmYear{2025}
\acmDOI{XXXXXXX.XXXXXXX}
\acmConference[HotCarbon’25]{The 4th Workshop on Sustainable Computer Systems}{July 10-11th, 2025}{Cambridge, MA}
\acmISBN{978-1-4503-XXXX-X/2025/06}

\tcbset{
  takeaway/.style={ enhanced,width=\hsize,left=0pt,right=0pt,top=0pt,bottom=0pt,colback=black!20!green!8!white,boxrule=1pt,colframe=black!30!green!30!white
  },
}
\newcounter{takeawaycounter}
\newcommand{\takeawaybox}[1]{
    \stepcounter{takeawaycounter}
    \begin{tcolorbox}[takeaway]
    \textbf{Takeaway~\arabic{takeawaycounter}:}#1
    \end{tcolorbox}
}

\tcbset{
  question/.style={ enhanced,width=\hsize,left=0pt,right=0pt,top=0pt,bottom=0pt,colback=black!20!red!8!white,boxrule=1pt,colframe=black!30!red!30!white
  },
}
\newcounter{questioncounter}

\colorlet{minenergy}{green!20}
\colorlet{lowenergy}{yellow!40}
\colorlet{midenergy}{orange!30}
\colorlet{highenergy}{red!30}
\colorlet{nodata}{gray!50}

\newcommand{\SYSTEM}{FABRIC}

\begin{document}

\input{tex/0-abstract}

\title{When Servers Meet Species: A Fab-to-Grave Lens on Computing's Biodiversity Impact}



\author{Tianyao Shi}
\orcid{0009-0006-6782-0144}
\affiliation{%
  \institution{Purdue University}
   \city{West Lafayette}
   \state{IN}
   \country{USA}}
\email{shi676@purdue.edu}

\author{Ritbik Kumar}
\orcid{0009-0003-8003-3418}
\affiliation{%
   \institution{Purdue University}
   \city{West Lafayette}
   \state{IN}
   \country{USA}}
\email{kumar684@purdue.edu}

\author{Inez Hua}
\orcid{0000-0003-4977-5758}
\affiliation{%
   \institution{Purdue University}
   \city{West Lafayette}
   \state{IN}
   \country{USA}}
\email{hua@purdue.edu}

\author{Yi Ding}
\orcid{0000-0003-2757-9182}
\affiliation{%
   \institution{Purdue University}
   \city{West Lafayette}
   \state{IN}
   \country{USA}}
\email{yiding@purdue.edu}

\thanks{This work is accepted to present at the 4th Workshop on Sustainable Computer Systems (HotCarbon'25). The proceeding version is published in ACM SIGEnergy Energy Informatics Review (EIR), Volume 5 Issue 2, July 2025.}

\maketitle

\input{tex/1-introduction}
\input{tex/2-background.tex}
\input{tex/3-modeling.tex}
\input{tex/4-evaluation}
\input{tex/5-conclusion.tex}

\bibliographystyle{ACM-Reference-Format}
\bibliography{reference}


\end{document}

%% file: tex/0-abstract.tex
\begin{abstract}

Biodiversity loss is a critical planetary boundary, yet its connection to computing remains largely unexamined. Prior sustainability efforts in computing have focused on carbon and water, overlooking biodiversity due to the lack of appropriate metrics and modeling frameworks. This paper presents the first end-to-end analysis of biodiversity impact from computing systems. We introduce two new metrics—Embodied Biodiversity Index (EBI) and Operational Biodiversity Index (OBI)—to quantify biodiversity impact across the lifecycle, and present FABRIC, a modeling framework that links computing workloads to biodiversity impacts. Our evaluation highlights the need to consider biodiversity alongside carbon and water in sustainable computing design and optimization.  The code is available at \url{https://github.com/TianyaoShi/FABRIC}.


\end{abstract}

%% file: tex/1-introduction.tex
\section{Introduction}


The environmental impact of computing has grown rapidly with the rise of AI, hyperscale datacenters, and semiconductor fabs~\cite{wu2022sustainable}. Prior work has focused on carbon~\cite{wu2022sustainable,radovanovic2022carbon,acun2023carbon,gupta2022act} and water~\cite{gnibga2024flexcooldc,Ali2024making,gupta2024dataset,jiang2025waterwise} dimensions of sustainability in computing. In contrast, biodiversity impact, a key endpoint in life cycle assessment (LCA) frameworks~\cite{curran2011toward}, remains largely unexamined in computing. From pollutant emissions during chip fabrication to toxic waste at end-of-life, computing systems generate stressors that contribute directly to species loss and ecosystem degradation~\cite{bonan2018climate}.

Despite its importance, biodiversity impact has remained absent from sustainable computing research. We identify two key barriers: \textbf{(1)} the lack of quantifiable metrics that can attribute biodiversity and ecosystem damage to specific computing activities; and \textbf{(2)} the lack of a practical modeling framework that links specific computing workload and system to concrete biodiversity impacts across their lifecycle---from manufacturing to operation to end-of-life.

To overcome these challenges, we introduce two new metrics: the Embodied Biodiversity Index (EBI) covering manufacturing, transportation, and end-of-life, and the \emph{Operational Biodiversity Index (OBI)} covering electricity use. They together provide a quantifiable basis for attributing biodiversity impact to computing systems.

Building on these metrics, we present \SYSTEM{} (\textbf{Fa}brication-to-grave \textbf{B}iodiversity \textbf{I}mpact \textbf{C}alculator), a new modeling framework that traces the fab-to-grave lifecycle of computing components (e.g., CPU, GPU, memory, storage) across manufacturing, transportation, use, and end-of-life stages. The key insight behind \SYSTEM{} is that biodiversity impact can be decomposed and attributed using a combination of LCA midpoint indicators (e.g., acidification, eutrophication, ecotoxicity). This enables us to directly connect real-world computing activities to measurable ecological consequences.

We evaluate \SYSTEM{} on 7 HPC workloads across 3 computing platforms with diverse devices. Key findings include:
\begin{itemize}
    \item Manufacturing and acidification dominate embodied biodiversity loss. Newer devices have lower impact per unit performance.
    \item Operational electricity outweighs manufacturing in biodiversity lifecycle impact under conservative assumptions.
    \item Low carbon does not always mean low biodiversity impact.
\end{itemize}
Our contributions are summarized as follows.
\begin{itemize}
    \item Conduct the first analysis of biodiversity impact in computing.
    \item Introduce two new metrics-EBI and OBI-to quantify biodiversity impact from manufacturing, transportation, end-of-life and use.
    \item Present \SYSTEM{}, the first modeling framework that connects computing workloads to lifecycle biodiversity impact.
    \item Evaluate biodiversity impact across devices, systems, workloads, and geographic deployment.
\end{itemize}


%% file: tex/2-background.tex
\section{Background}

In this section, we review the LCA framework for assessing biodiversity impacts of computing and define the scope of our study. We focus on two hierarchical levels: \emph{mid-point} impacts, which capture specific pollutant-driven mechanisms, and \emph{end-point} impact, which quantifies resulting damage to ecosystems and species.


\subsection{Mid-point Impact Categories}

In this paper, we focus on three midpoint impacts: acidification, eutrophication, and ecotoxicity; as sustainability reports from key vendors disclose emission data that directly map to these impacts. Other midpoints like ozone depletion and smog formation are excluded due to limited data \cite{hetherington2014use} and lower contribution to endpoint damage \cite{van2024revisiting}.

\noindent\textbf{Acidification (AP)}, measured in kilograms of SO\textsubscript{2} equivalents, reflects how much emitted gases--- SO\textsubscript{2}, NO\textsubscript{x}, and NH\textsubscript{3}---can lower environmental pH. These gases form acid rain, which leaches nutrients from soil and harms plants and aquatic life. In computing, AP largely results from SO\textsubscript{2} and NO\textsubscript{x} emissions from fossil fuel combustion in chip manufacturing and datacenter electricity use.

\noindent\textbf{Eutrophication (EP)}, reported in kilograms of phosphate equivalents (kg PO\textsubscript{4}\textsuperscript{3–} eq), measures how excess nitrogen and phosphorus enrich ecosystems \cite{smith1999eutrophication}, often leading to oxygen-depleting algal blooms. In computing, EP stems from wastewater in chip production, chemical slurries used in processing, and nitrate/ammonia emissions from power generation.

\noindent\textbf{Freshwater ecotoxicity potential (FETP)}, expressed in comparative toxic units for ecosystems (CTU\textsubscript{e}), captures the toxic impact of chemicals on aquatic ecosystems \cite{rosenbaum2008usetox}. In computing, key sources include heavy metals (e.g., Cu, Ni), photoresist residues, fluorinated surfactants, and waste from both fabs and material extraction.



\subsection{End-point Impact Metric}

End-point biodiversity impact in \textbf{species·yr}, which quantifies the expected fraction of species locally lost per year due to cumulative environmental stressors. For example, a value of $1 \times 10^{-4}$ species·yr means one ten-thousandth of a species is statistically lost in a region over a year. To convert acidification, eutrophication, and ecotoxicity impacts into species·yr, we use the ReCiPe 2016 ecosystem damage model \cite{huijbregts2016recipe} --- chosen over ILCD 2011 \cite{chomkhamsri2011international} and IMPACT World+ \cite{bulle2019impact} for its comprehensive treatment of ecological impacts. This unified metric enables us to compare different environmental stressors and assess the biodiversity risks embedded in computing infrastructure.


\subsection{Scope Discussion}

In this paper, we use a \emph{fab-to-grave} system boundary that includes the manufacturing, transportation, use, and end-of-life stages (\Cref{fig:diagram}). We exclude upstream mining and material refinement despite their significance, due to fragmented and opaque global supply chains that limit reliable impact tracing. Following the Dell R740 LCA report~\cite{Busa2019_DellR740_LCA}, we assume packaging uses fully recycled materials with negligible biodiversity impact. We focus on five key computing components---CPU, GPU, DRAM, SSD, and HDD---which account for most semiconductor wafer throughput, operational energy use, and pollutant emissions driving AP, EP, and FETP. As such, our reported results reflect a lower bound on biodiversity impact, and will become more complete as public data expands. 



%% file: tex/3-modeling.tex
\begingroup
\begin{figure*}
\setlength{\textfloatsep}{0pt}
\captionsetup{aboveskip=2pt, skip=3pt}
    \centering
    \includegraphics[width=0.99\textwidth]{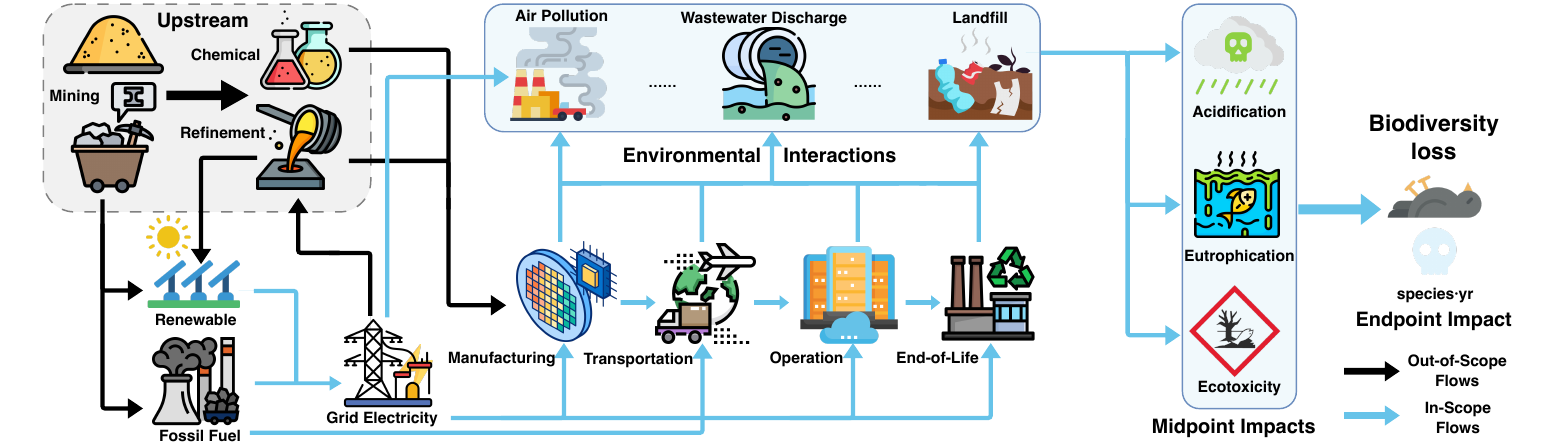}
    \caption{The \textbf{Fa}brication-to-grave \textbf{B}iodiversity \textbf{I}mpact \textbf{C}alculator \textbf{(FABRIC)} modeling framework.}
    \label{fig:diagram}
\end{figure*}
\endgroup

\section{The \SYSTEM{} Modeling Framework}\label{sec:model}


\Cref{fig:diagram} shows the workflow of \SYSTEM{}. We first introduce the two newly defined metrics for quantifying biodiversity impact, and then explain their application across different lifecycle stages.

\subsection{Definitions of EBI and OBI}


The total biodiversity impact is the summation of EBI and OBI. Next, we introduce them separately.

\textbf{Embodied Biodiversity Index (EBI)} refers to the biodiversity damage incurred once over the physical lifecycle of a device that covers the manufacturing (\textsf{Mfg}), transportation (\textsf{Trans}), and end-of-life (\textsf{EoL}) stages. The device harms the environment in two key ways: (1) directly releasing toxic substances through leaks, discharge, or e-waste; and (2) indirectly causing emissions by driving fossil fuel use for transport and electricity generation.
The emissions in each stage $l\!\in\!\{\textsf{Mfg},\textsf{Trans},\textsf{EoL}\}$ of a device $d$ lead to multifaceted midpoint impacts $M_{c,l}(d)$, where $c\!\in \!\mathcal{C}\!=\!\{\textsf{AP},\textsf{EP},\textsf{FETP}\}$ is the midpoint impact category. The EBI of a device is calculated by summing impacts across stages and impact categories:

\begin{equation}
   B_{\text{emb}}(d) \;=\;
   \sum_{c,l}
    M_{c,l}(d)\,
    \Phi_{c},
    \label{eq:embodied}
\end{equation}
where $\Phi_{c}$ is the midpoint-to-endpoint conversion factor for $c$. 

Similar with the definition of embodied carbon, EBI of a specific workload is proportional to the execution time amortized over device lifetime. Let $LT_d$ be the lifetime of device $d$,  the EBI of a specific workload $w$ with execution time $t_{w}$ is given by

\begin{equation}
    B_{\text{emb}}(w) \;=\; \sum_{d\in\mathcal{D}} \frac{t_w}{LT_d} B_{\text{emb}}(d),
\end{equation}
where $\mathcal{D}$ is the set of all devices involved in the execution.

\textbf{Operational Biodiversity Index (OBI)} is the biodiversity damage incurred from generating the electricity needed for the operational use of computing devices.
\begin{equation}
   B_{\text{op}}(d) \;=\;
   \sum_{c}
    M_{c,\textsf{Use}}(d)\,
    \Phi_{c}.
\end{equation}

Similar with operational carbon, the task-specific OBI of $w$ is proportional to its energy consumption $E_{\mathrm{el}}(w)$: 
\begin{align}
   B_{\text{op}}(w) \;&=\;
    \sum_{c}M_{c,\textsf{Use}}(w)
    \Phi_c, \\
    M_{c,\textsf{Use}}(w)\;&=\;\sum_{k}
    E_{\mathrm{el}}(w)  F_{\text{el},k,r}\,
    \Gamma^c_{k}\,.
    \label{eq:operational}
\end{align}
\Cref{eq:operational} reveals details on how to compute the midpoint impacts by summing grid-related emission loads: 
$k\!\in\!\{\text{SO}_2,\text{NO}_x,\text{NH}_3\}$ is the associated pollutant species, $F_{\text{el},k,r}$ is the emission factor for region $r$ (in g/kWh), and $ \Gamma^c_{k}$ is the \emph{characterization factor} to convert the pollutant $k$ into the reference unit for impact category $c$. For example, 1 kg NH$_3$ is 0.93 kg SO$_{2\text{ eq}}$ for AP \cite{BIOIS2005_ELV_Annex5}.

Generally, the midpoint impact of a device $d$'s lifecycle stage $l$ for impact category $c$ is given by
\begin{equation}
    M_{c,l}(d) = \sum_{j,k} E_{j}(d)\,F_{j,k,r}\Gamma_{k}^{c},
    \label{eq:midpoint}
\end{equation}
where \(E_{j}\) denotes the measurable throughput of process \(j\) (e.g., kWh for electricity use, wafer‐mask-layers for logic chips manufacturing, or ton-kilometers for transportation) and \(F_{j,k,r}\) is the corresponding per-unit emission factor of pollutant species $k$ in region $r$.




\textbf{Summary.} Since the biodiversity impact from electricity use falls under OBI and has been thoroughly covered above, we will focus next on modeling EBI---specifically from manufacturing, transportation, and end-of-life stages.


\subsection{EBI from Manufacturing} 


Different devices have distinct manufacturing processes and have varying levels of data availability, so we calculate their EBIs separately. 
As implied by \Cref{eq:midpoint}, the key is to determine the values of process throughput $E_j$ (i.e., how many production units it takes to manufacture a device) and the per-unit emission factors \(F_{j,k,r}\) for each type of computing device.

\subsubsection{Logic devices (CPU, GPU core)}
This is an example where both $E_j$ and \(F_{j,k,r}\) can be easily acquired:
\begin{align*}
 E_{\textsf{Mfg}}(d) = \frac{\text{Die Size}}{\text{Wafer Area}\cdot\text{Yield}} \cdot \#\text{wafer-mask-layers},
\end{align*}
where \#wafer-mask-layers is  67, 78, 87, and 81 layers for 14, 10, 7, and 5nm node, respectively \cite{Schor2020_TSMC_5nm}. For \(F_{j,k,r}\), the annual per-production-unit air pollutant emission, wastewater discharge, and electricity consumption is available in TSMC sustainbility reports \cite{TSMC_ESG_Documents}.

\subsubsection{DRAM and SSD}
This is an example where only fab-level annual emission data is available but the fab produces multiple types of devices.
According to ISO 14044 \cite{finkbeiner2006new}, we use economic-value-based allocation to obtain per-unit emission factors.
Let \(p\!\in\!\{\textsf{DRAM},\textsf{SSD}\}\) be the product class of $d$:  
\begin{align*}
  E_{\textsf{Mfg}}(d) &= \frac{\text{Capacity}_d}{\text{Wafer Area}\cdot\text{Bit Density}_p/8}\cdot\frac{1}{\text{Yield}}, \\
  F_{\textsf{Mfg},k,r}(d) &= \frac{\text{Revenue Share}_p \cdot \text{Annual Total Emission}_k}{\text{Wafer Production Capacity}_p},
\end{align*}
where $E_{\textsf{Mfg}}(d)$ means expected number of wafers to produce a single device $d$ and $F_{\textsf{Msf},k,r}(d)$ means per-wafer emission of pollutant species $k$.
The emission from electricity use is derived similarly. 
For GPUs, the attached high-bandwidth memory (HBM) adopts the \textsf{DRAM} formula and the total EBI of GPU adds the two parts up.

\subsubsection{HDD} 
Cradle-to-grave LCA studies~\cite{Seagate2016_Makara_HDD_LCA,jin2020life} already provide per-unit midpoint scores. We extrapolate these scores for the missing years in our study scope by assuming similar overall reductions as seen in carbon emissions~\cite{Seagate_ESG_Product_Sustainability}. The values are directed used as \(M_{c,l}(d_\textsf{HDD})\). Importantly, the broader system boundaries in LCAs only reinforce the conservativeness of our lower-bound estimates.


\subsection{EBI from Transportation}

The process throughput of transporting is
\begin{align*}
    E_{\textsf{Trans}.g}(d) = \sum_m \text{Shipping Mass}_d \cdot \text{Distance}_g, 
\end{align*}
where $g$ is the transport mode (e.g., truck, ship). The mode‑specific emission factors $F_{\textsf{Trans}.g,k,r}$ are available in Ecoinvent database \cite{wernet2016ecoinvent}.
Note that emission factors vary over time due to stricter regulations on maritime fuel sulfur content and truck emissions \cite{grigoratos2019real}.
For simplicity, we assume a fixed transport profile: 200km by truck (100km from the fab to the port and 100km from the port to the datacenter) and 14,000km by ship (East Asia to the U.S. west coast).

\subsection{EBI from End‑of‑Life (EoL)}
The end-of-life (EOL) stage requires a different modeling approach due to its complexity.
Discarded servers and storage devices typically follow three disposal pathways: \emph{recycling} (disassembly, metal shredding/smelting, and plastics recovery), \emph{incineration}, and controlled \emph{landfill}, with ratios denoted as $R$, $I$, $L$, respectively.  
Given the mass $m_d$, the midpoint impact is
\begin{equation}
  M_{c,\textsf{EoL}}(d) \;=\; 
  m_d \bigl[\, R\,F^{c}_{\text{rec}} \;+\; 
             I\bigl(F^{c}_{\text{inc}} + p_{\text{ash}}F^{c}_{\text{ash}}\bigr) \;+\; 
             L\,F^{c}_{\text{land}} \bigr],
  \label{eq:eol}
\end{equation}
where \(R+I+L=1\), \(p_{\text{ash}}\) is the kilograms of bottom ash per kilogram of feed incinerated, and \(F^{c}_{\bullet}\) is the net impact for each pathway.   
However, direct \(F^{c}_{\bullet}\) values for mixed electronic waste are rarely reported, making detailed modeling intractable. Therefore, we use precise EoL impacts from the latest Fairphone 5 LCA study~\cite{Sanchez2024_Fairphone5_LCA} as proxies for the recycling pathway~\Cref{eq:midpoint}:
\begin{align*}
   E_{\textsf{EOL}}(d)=m_d, \qquad \sum_{k}F_{\textsf{EOL},k}CF_k^c = \frac{M_{c,\textsf{EOL}}(\text{Fairphone 5})}{m_\text{Fairphone 5}}. 
\end{align*}
We assume each category's impact scales linearly with device mass, as the bills of materials of computing devices distribute similar ratios to metal, plastic, etc. Then, we apply these estimates uniformly across all devices until more specific data become available.



%% file: tex/4-evaluation.tex
\section{Evaluation}\label{sec:results}

We examine the following research questions (RQs):
\begin{enumerate}[label=\textbf{RQ\arabic*:}]
\item Which lifecycle stage contributes most to computing’s biodiversity impact? (\Cref{subsec:rq1})
\item How do biodiversity impacts vary across HPC workloads and system platforms? (\Cref{subsec:rq2})
\item How does deployment location affect computing’s biodiversity impacts? (\Cref{subsec:rq3})
\end{enumerate}

\subsection{Evaluation Methodology}

\begin{table}[t]
\centering
\caption{Individual devices studied in this paper.}
\label{tab:device_specs}
\vspace{-1em}
\resizebox{\linewidth}{!}{%
\begin{tabular}{lcccc}
\toprule
\textbf{CPU} & \textbf{Year} & \textbf{Node} & \textbf{Silicon Area} & \textbf{Cores} \\
\midrule
AMD EPYC 7B12  & 2019 & 7/14 nm & 1\,008 mm$^2$ & 64 \\
AMD EPYC 7443  & 2021 & 7/12 nm &   740 mm$^2$ & 24  \\
AMD EPYC 7B13  & 2021 & 7/12 nm & 1\,064 mm$^2$ & 64 \\
AMD EPYC 9B14  & 2023 & 5/6 nm & 1\,261 mm$^2$ & 96 \\
\midrule
\textbf{GPU} & \textbf{Year} & \textbf{Node} & \textbf{Silicon Area} & \textbf{VRAM} \\
\midrule
NVIDIA T4          & 2018 & 12 nm  &   545 mm$^2$ & 16 GB\\
NVIDIA V100        & 2017 & 12 nm &   815 mm$^2$ & 16 GB\\
NVIDIA L40         & 2022 & 5 nm  &   609 mm$^2$ & 48 GB\\
NVIDIA A100 & 2020 & 7 nm  &   826 mm$^2$ & 40 GB \\
NVIDIA H100 & 2023 & 5 nm  &   814 mm$^2$ & 80 GB\\
\midrule
\textbf{Device} & \textbf{Vendor} & \textbf{Type} & \textbf{Year} & \textbf{Capacity}   \\
\midrule
DDR4 RDIMM & SK Hynix & DRAM  & 2020 & 64 GB      \\
PE8111  & SK Hynix       & SSD   & 2022 & 15.36 TB   \\
Exos X20  & Seagate      & HDD   & 2023 & 20 TB    \\
\bottomrule
\end{tabular}}
\end{table}

\begin{table*}[t]
  \centering
  \caption{Three computing systems evaluated in this paper.}
  \label{tab:system_specs}
  \vspace{-1em}
  \resizebox{\textwidth}{!}{
    \begin{tabular}{lcccccc}
      \toprule
      \textbf{System} & \textbf{Main HW Year} &
      \textbf{CPUs (model $\times$ count)} &
      \textbf{GPUs (model $\times$ count)} &
      \textbf{Total DRAM} &
      \textbf{Storage Capacity} \\
      \midrule
      Local testbed server &
      2022 &
      EPYC 7443 \(\times\) 1 &
      L40 \(\times\) 4 &
      520 GB &
      2 TB SSD + 20 TB HDD \\[2pt]
      Gautschi$^*$ \cite{rcac2025_gautschi}  &
      2023 &
      EPYC 9B14 \(\times\) 442 &
      H100 \(\times\) 160 + L40 \(\times\) 12 &
      158 TB &
      1.5 PB SSD + 2.5 PB HDD\\[2pt]
      Perlmutter \cite{nersc2023_perlmutter} &
      2021 &
      EPYC 7B13 \(\times\) 4 864 &
      A100 80 GB \(\times\) 1 024 + A100 40 GB \(\times\) 6 144 &
      1 984 TB &
      44 PB SSD \\
      \bottomrule
    \end{tabular}}
\end{table*}


\noindent\textbf{Device and system scopes.} To analyze device-level trends over time, \Cref{tab:device_specs} lists the specifications of individual devices across a range of generations and manufacturing technologies used in our \SYSTEM{} analysis. 
To examine how biodiversity impacts scale from a single node to large-scale systems, we evaluate three computing setups shown in \Cref{tab:system_specs}: a \emph{local testbed server} representing a typical edge or lab workstation, \emph{Gautschi}$^*$ representing a 2023-era community cluster in academia~\cite{rcac2025_gautschi}\footnote{We replaced 40 Intel CPUs with comparable AMD EPYC 9B14 processors to enable direct comparison.}, and \emph{Perlmutter} representing a petascale supercomputer with mixed CPU/GPU architecture.

\noindent\textbf{Workloads.} We evaluate seven HPC workloads from the Phoronix Test Suite~\cite{phoronix2025}, spanning lossless compression/decompression, scientific/analytic kernels, media encoding, and software compilation (see x-axis in \Cref{fig:res_hpc}). These workloads represent a broad range of compute, memory, and storage demands, capturing both legacy and emerging biodiversity stressors.

\noindent\textbf{Data sources.}  
We make our best effort to ensure the credibility of publicly accessible sources used in this study by selecting leading vendors' sustainability reports, authoritative agency documents, and highly-recognized peer-review studies.
The values for midpoint characterization factors \(\Gamma_{k}^{c}\) are determined based on TRACI documentation~\cite{bare2011traci,traciv22} from the United States Environmental Protection Agency (EPA). 
The midpoint-to-endpoint conversion factors $\Phi_c$ are based on ReCiPe 2016 Hierarchist framework~\cite{huijbregts2016recipe}.
Logic-device data derive from TSMC's sustainability reports \cite{TSMC_ESG_Documents}, and memory data from SK Hynix \cite{SKhynix_Sustainability_Reports}; revenue shares come from audited financials \cite{SKhynix2025_Audit_Report}, wafer capacity from TrendForce \cite{trendforce2025_nandfab}, and bit densities from IEDM paper \cite{jones2023modeling} and TechInsights \cite{techinsights2024_hbmtrend,techinsights2019_iedm18,singer2019_dramnand}.
All datasets are limited to the 2016–2023 window, aligning with the temporal scope of our study.
The yield of devices is assumed to be a constant 0.875 as in prior work~\cite{gupta2022act}.
Electricity consumption is characterized with SO\(_x\)/NO\(_x\)/NH$_3$ factors from EPA eGRID \cite{epa2025_egrid2023rev1}, EDGAR \cite{jrc2024_edgar2024}, and East-Asian utilities' (Taipower, KEPCO) sustainability reports \cite{taipower2025_csrdownload,kepco2025_sustainability}.
Although emission factors can differ across vendors due to varied manufacturing processes, TSMC and SK Hynix together dominate advanced logic and memory production, providing a reasonable reference for the sector. Any cross-vendor variance is therefore unlikely to overturn our trend-level conclusions.

\subsection{Biodiversity Impact Breakdown}\label{subsec:rq1}
\begingroup
\begin{figure}
\setlength{\textfloatsep}{-10pt}
\setlength{\floatsep}{-5pt}
\captionsetup{aboveskip=2pt, skip=5pt}
    \centering
    \includegraphics[width=0.99\linewidth]{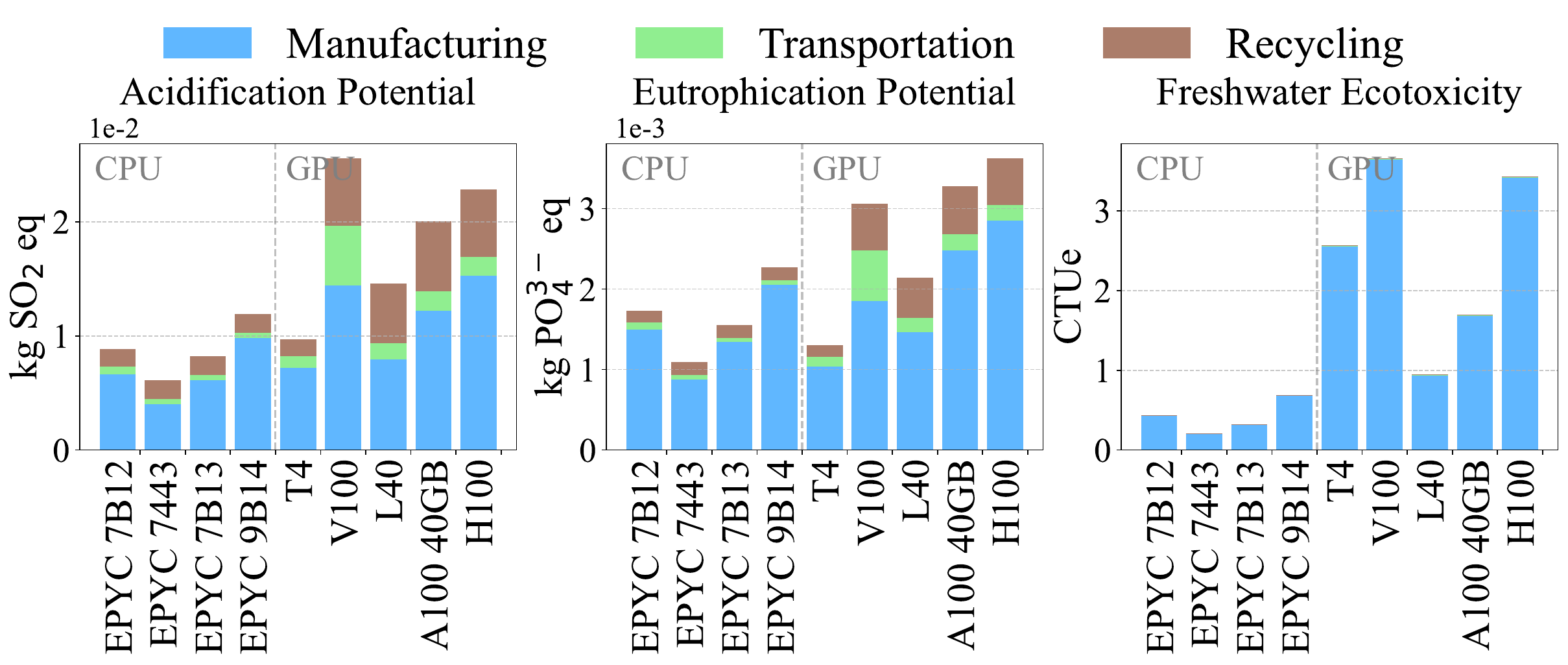}
    \caption{Midpoint EBI breakdown by lifecycle stages for CPUs and GPUs modeled.}
    \label{fig:mebi_logic}
\end{figure}

\begin{figure}
\captionsetup{aboveskip=2pt, skip=5pt}
    \centering
    \includegraphics[width=0.99\linewidth]{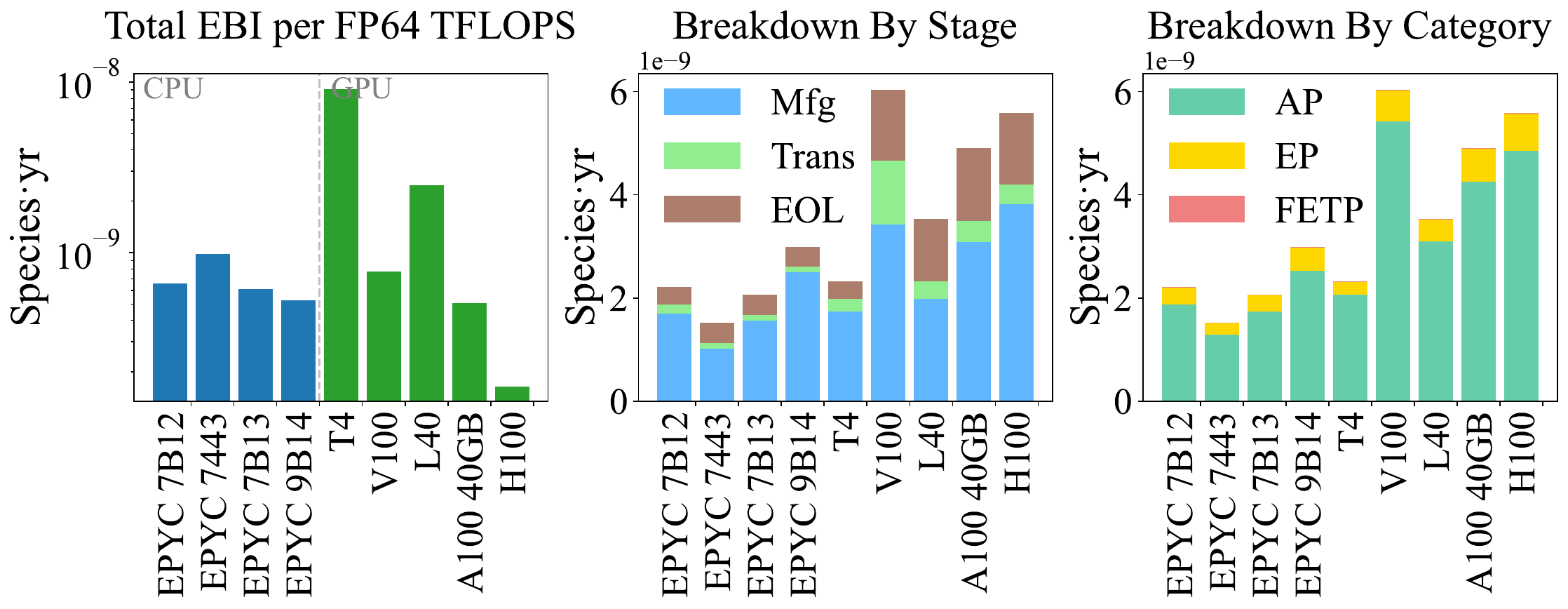}
    \caption{Endpoint EBI analyses for CPUs and GPUs modeled. Left: endpoint EBI per FP64 Teraflops. Middle: EBI contribution from different lifecycle stages. Right: EBI contributions from different midpoint impact categories.}
    \label{fig:ebi_logic}
\end{figure}
\endgroup

\noindent\textbf{Mid-/End-point EBI analysis.}  
\Cref{fig:mebi_logic} shows that \emph{manufacturing} overwhelmingly dominates device-level acidification (AP), eutrophication (EP), and freshwater-ecotoxicity (FETP), contributing 78–92\% of total mid-point impacts. In contrast, transportation and end-of-life (EOL) together rarely exceed 20\%.  GPUs generally have higher absolute impacts than CPUs—e.g.\ the 12nm \textsc{V100} emits nearly \(0.020\)\,kg SO\(_2\) eq of AP, more than twice that of the 7nm \textsc{EPYC 7B12}—yet node scaling alone cannot offset larger package masses, as the 5nm, 700W \textsc{H100} still tops the AP chart. When normalized by peak FP64 throughput (\Cref{fig:ebi_logic}, left), this trend reverses: the legacy double-precision GPUs (\textsc{V100}) incur \(\sim\!1.1\times10^{-8}\)\,species·yr TFLOPS\(^{-1}\), roughly 15$\times$ higher than the tensor-dense \textsc{H100} or modern CPUs---highlighting how architectural improvements and silicon scaling reduce embodied biodiversity damage. \Cref{fig:ebi_logic} shows that manufacturing remains the largest contributor to endpoint biodiversity impact (55–75\%), with acidification as the dominant pathway (60–85\%), followed by EP and FETP. A similar trend holds for memory and storage (\Cref{fig:ebi_mem_storage}): manufacturing dominates due to their minimal shipping mass per GB (< 1 gram), and capacity-normalized impacts drop nearly tenfold from 2016 to 2023.


\takeawaybox{
Manufacturing and acidification drives embodied biodiversity impact for computing devices. While absolute damage increases with newer generations, species·yr per TFLOP decreases, highlighting that newer, more efficient devices are environmentally preferable on a per-performance basis.
}


\begingroup
\begin{figure}
\setlength{\textfloatsep}{-10pt}
\setlength{\floatsep}{-5pt}
\captionsetup{aboveskip=2pt, skip=3pt}
    \centering
    \includegraphics[width=0.99\linewidth]{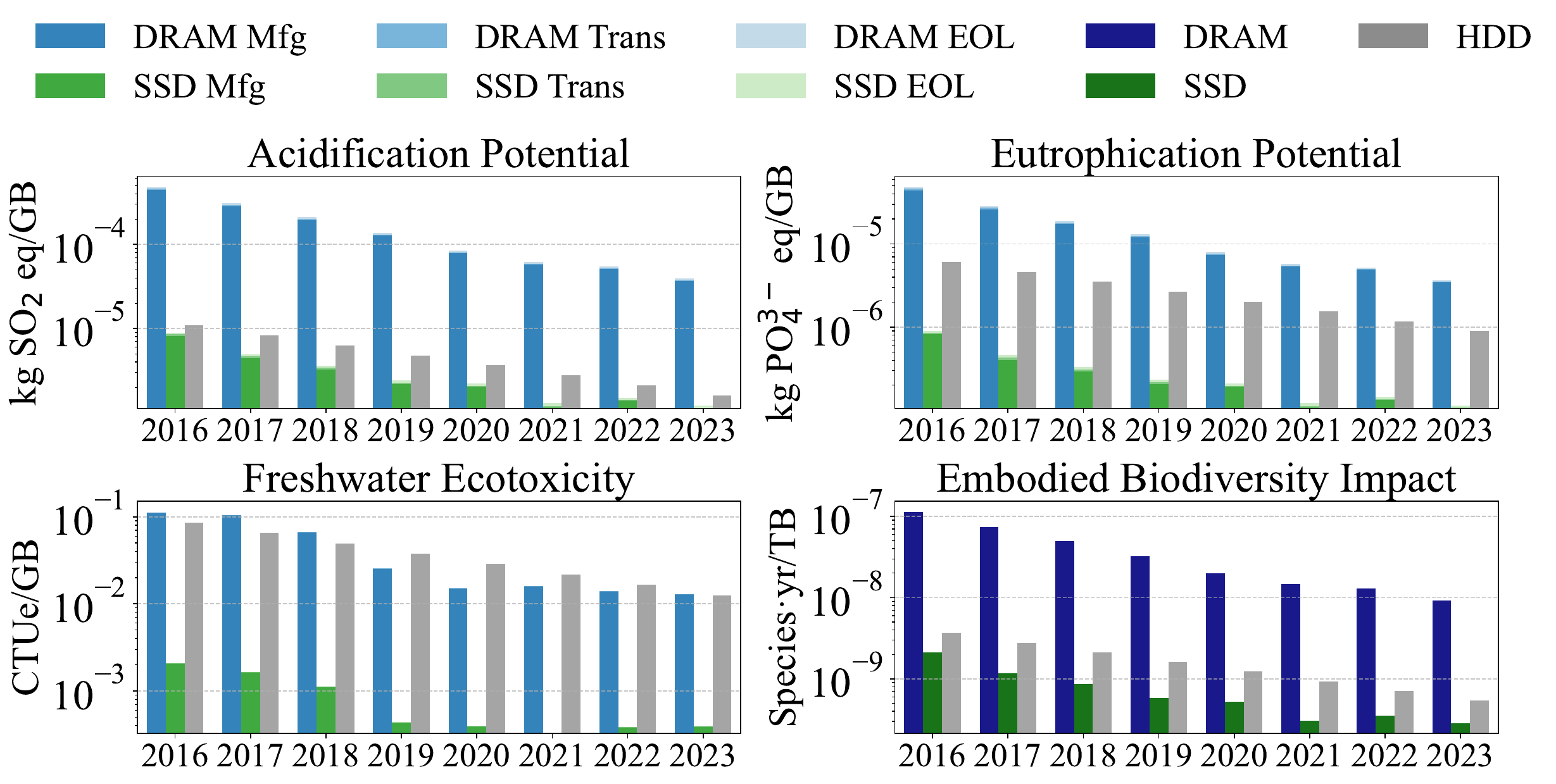}
    \caption{Mid-/End-point EBI for memory and storage.}
    \label{fig:ebi_mem_storage}
\end{figure}

\begin{figure}
\captionsetup{aboveskip=2pt, skip=3pt}
    \centering
    \includegraphics[width=0.99\linewidth]{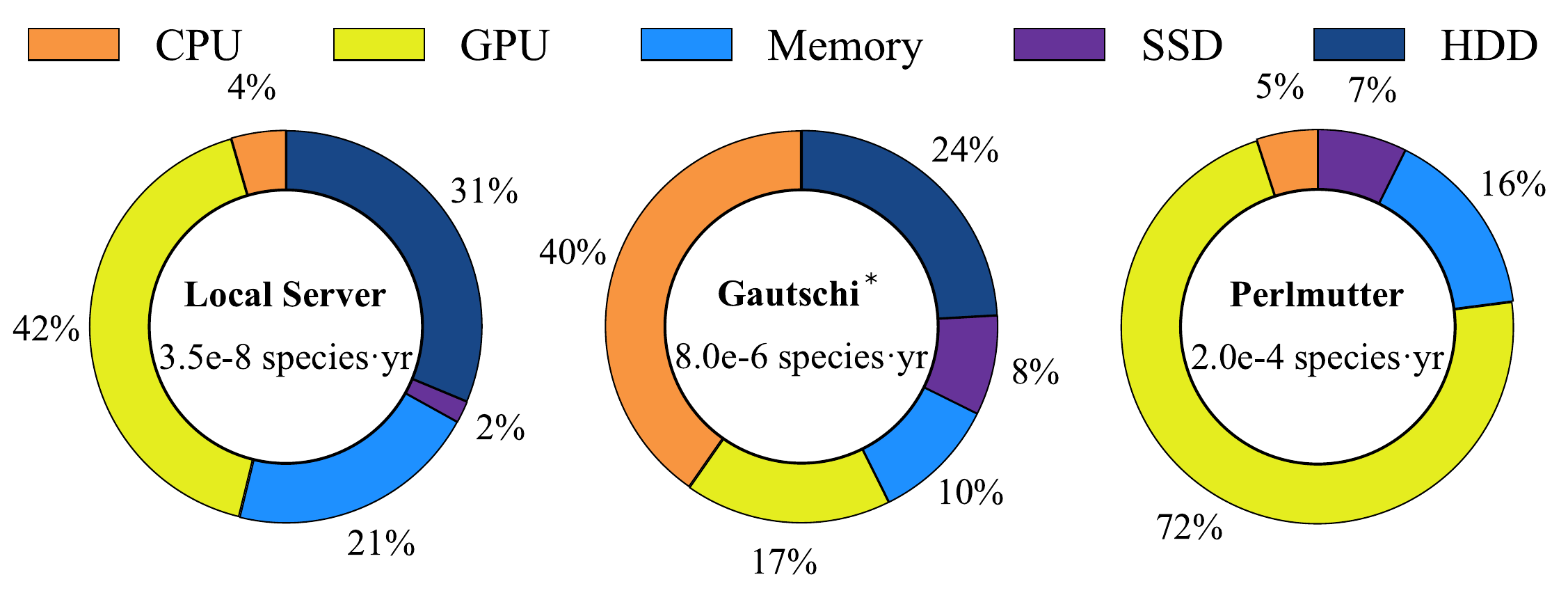}
    \caption{EBI contributions from different components in three studied systems.}
    \label{fig:sys_comp}
\end{figure}
\endgroup

\noindent\textbf{System-level EBI contributions.}  
\Cref{fig:sys_comp} shows results for full systems. A single–socket \emph{local server} carries an EBI of only \(3.5\times10^{-8}\)\,species·yr, whereas the GPU-heavy \emph{Perlmutter} reaches \(2.0\times10^{-4}\)\,species·yr—\,five orders of magnitude higher. These differences reflect system composition. GPUs dominate impact when they dominate hardware: contributing 72\% of Perlmutter's total but only 17\% in CPU-heavy \emph{Gautschi}$^*$, where CPUs and HDDs (a by-product of the large file system) account for 40.3\% and 24.3\%, respectively. Memory contributes 10–31\%, while SSDs remain below 8\%.

Perlmutter highlights the long-term impact of operations. At 70\% load and using U.S. grid average SO\textsubscript{2} and NO\textsubscript{x} emission factors, its annual electricity use of 27.4GWh adds \(\mathbf{2.51\times10^{-3}}\)\,species·yr—about 60$\times$ higher than its annualized EBI. This makes OBI dominate its lifecycle footprint.
If 400 Perlmutter-class systems are active globally, the sector causes \(\gtrsim\!1\)\,species·yr of biodiversity impact per year, with a footprint that will grow proportionally with future scale.


\begingroup
\begin{figure}
\setlength{\textfloatsep}{-10pt}
\setlength{\floatsep}{-5pt}
    \captionsetup{aboveskip=2pt, skip=2pt}
    \centering
    \includegraphics[width=0.99\linewidth]{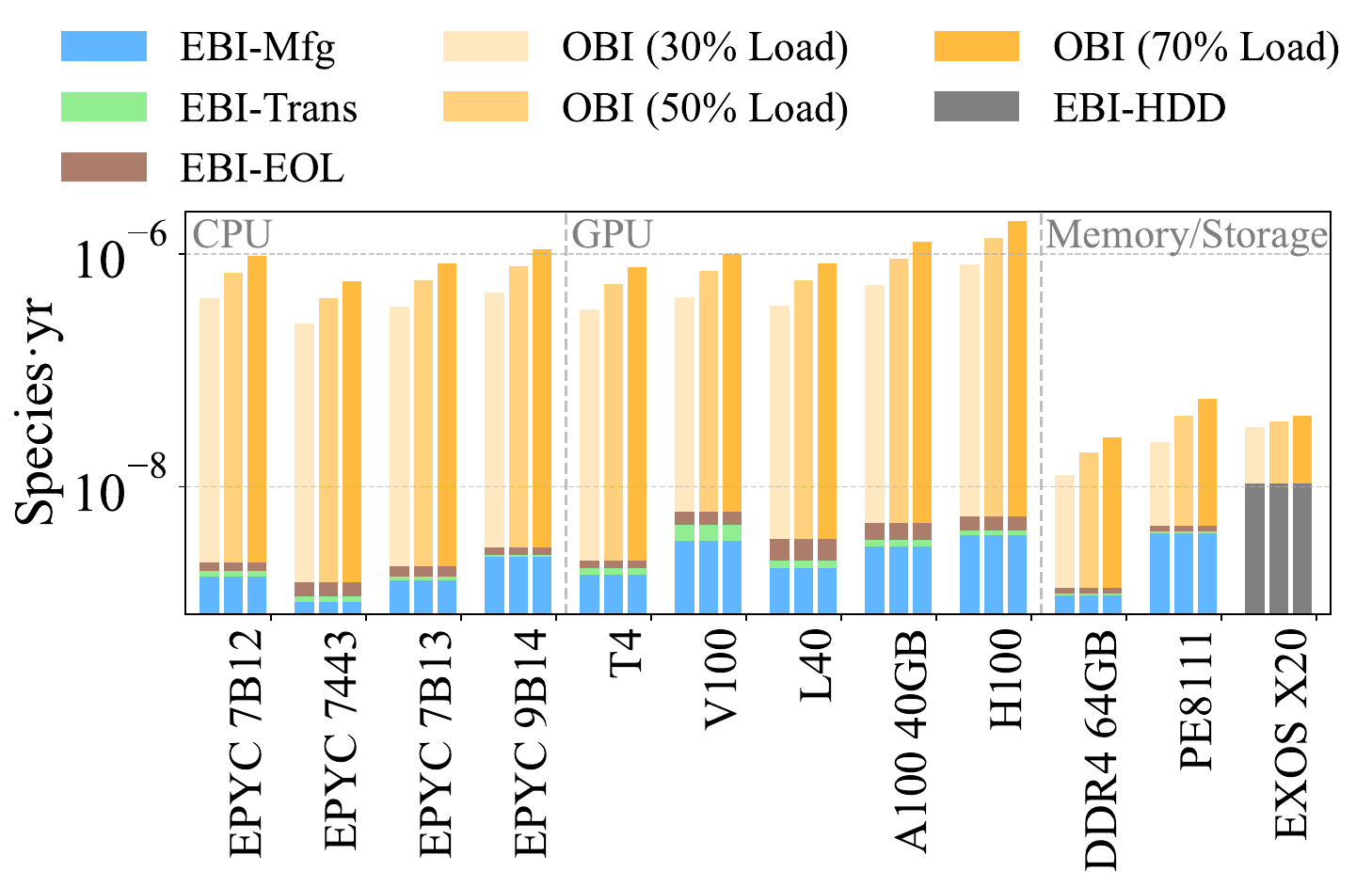}
    \caption{A lifecycle analysis for individual devices modeled.}
    \label{fig:logic_ebi+obi}
\end{figure}
\endgroup

\noindent\textbf{OBI dominance.} \Cref{fig:logic_ebi+obi} presents a full lifecycle view by stacking the embodied impacts (EBI) with annual operational impacts (OBI) across three representative duty cycles (30\%, 50\%, 70\%). 
Assuming U.S. grid average emission factors, electricity use pushes all devices above their embodied impact for all cases. At 70\% load, OBI outweighs EBI by nearly 100$\times$ for high-power CPUs and GPUs.
Although the mass-heavy 20TB HDD shows the largest EBI across all devices, its OBI is modest, occupying less than 10\% total impact of logic devices.  
This suggests that improving energy efficiency and adopting cleaner electricity are the most effective ways to lower a system's total biodiversity impact.

\takeawaybox{
When using conservative embodied estimates that exclude raw material extraction and refining, and assuming average U.S. grid emissions, operational electricity—not manufacturing—dominates a device’s lifecycle biodiversity impact.
}



\subsection{Workload-Level Impact Assessment}\label{subsec:rq2}

We test seven HPC workloads on three platforms: (1) \emph{Google Cloud Platform} (GCP) \textsf{n2d-standard-64} with 32 physical cores (half of EPYC 7B12) \cite{gcp_general_machine}, (2) the local server in \Cref{tab:system_specs}, and (3) GCP \textsf{c3d-standard-60} with 30 physical cores (5/16 of EPYC 9B14), representing a generational progression of HPC systems every 2–3 years.

To ensure fair comparison, all platforms are assumed to have 256GB memory and 100GB HDD and located in the U.S. Midwest. Since all workloads are CPU-centric, we do not consider impacts from additional devices like GPUs on the local server. 
We normalize throughput, energy consumption per unit throughput, and total biodiversity impact per unit throughput using the oldest GCP \textsf{n2d} as the reference point. Results are shown in \Cref{fig:res_hpc}. While performance increases—up to 55\% with the newest VM—energy efficiency tells a clearer story: the local server consumes 1.3–3.2$\times$ more energy per unit work, likely due to less optimized power management. Consequently, biodiversity impact follows the same trend: the 5nm cloud VM reduces impact to 55–75\% of baseline, while the local server increases it by 30–140\%, nullifying performance gains.

\begingroup
\begin{figure}
\setlength{\textfloatsep}{-10pt}
\setlength{\floatsep}{-5pt}
    \captionsetup{aboveskip=2pt, skip=2pt}
    \centering
    \includegraphics[width=0.99\linewidth]{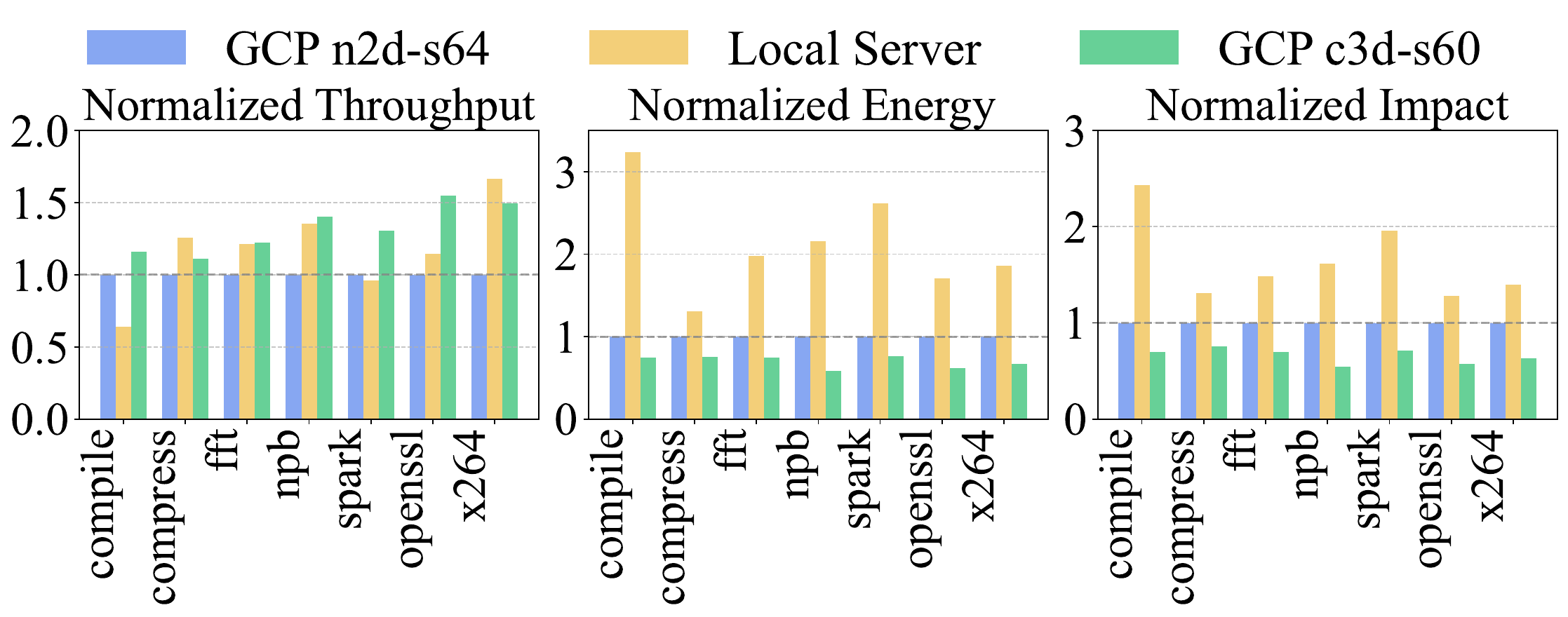}
    \caption{Normalized throughput, energy consumption, and total biodiversity impacts for HPC workloads.}
    \label{fig:res_hpc}
\end{figure}
\endgroup

\takeawaybox{ 
With the same memory and storage, newer energy-efficient CPUs in cloud environments reduce biodiversity impact of per unit of work by half, while a local server can double it—even if performance appears comparable. This indicates that datacenter-level power management may matter more than chip technology.
}


\subsection{Geospatial Sensitivity}\label{subsec:rq3}

\begingroup
\begin{figure}
\setlength{\textfloatsep}{-10pt}
\setlength{\floatsep}{-5pt}
    \captionsetup{aboveskip=2pt, skip=2pt}
    \centering
    \includegraphics[width=0.99\linewidth]{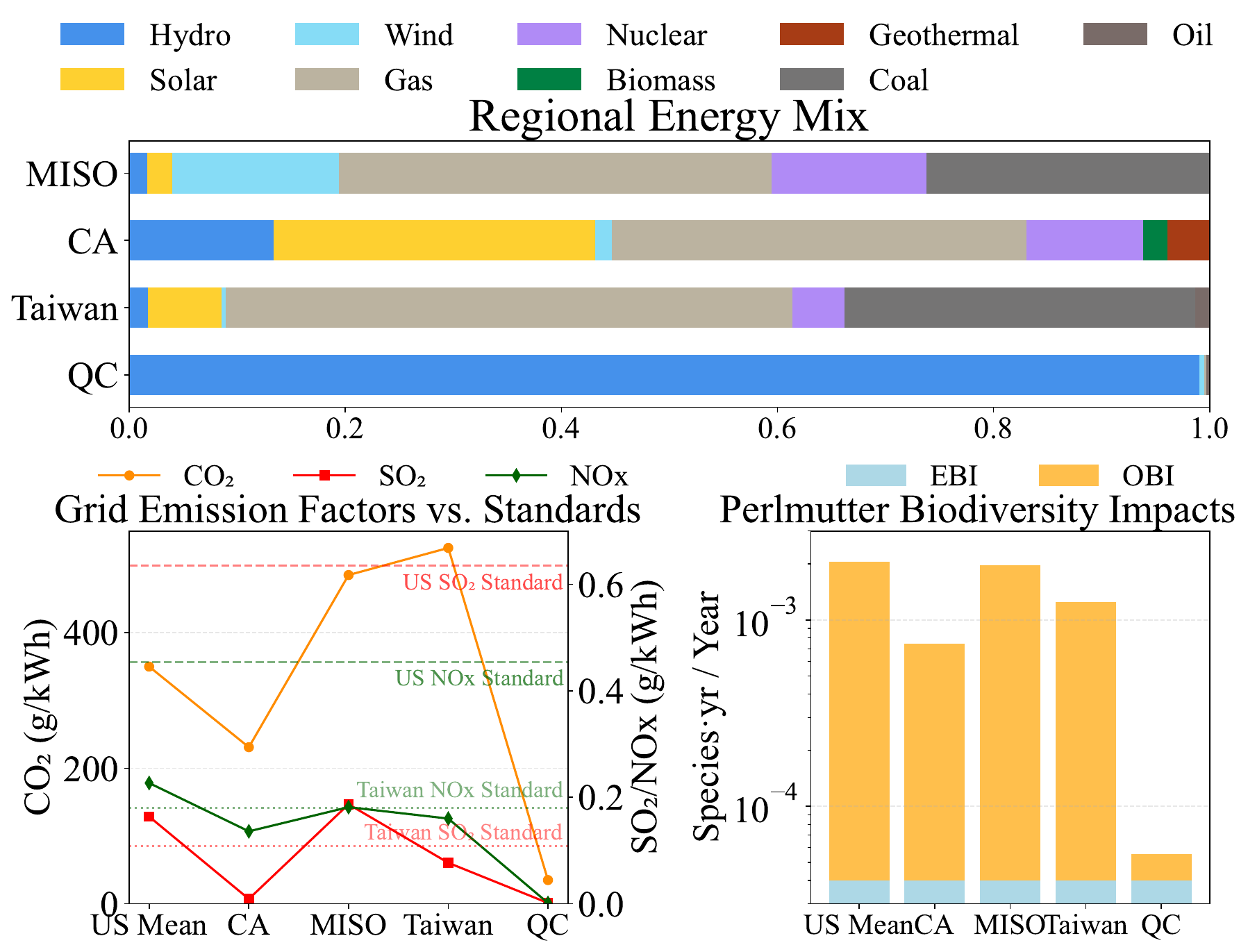}
    \caption{Regional Grid Emissions vs. Biodiversity Impacts.}
    \label{fig:spatial}
\end{figure}
\endgroup

Because emission factors vary by location, we perform a geospatial sensitivity analysis by hypothetically placing the \textit{Perlmutter} system in four distinct electricity markets: California (CA), the Midwest (MISO), Taiwan, and hydro-rich Québec (QC).
\Cref{fig:spatial} shows the results.
The findings highlight that low-carbon does not always mean low biodiversity impact. For example, CA's solar-heavy grid and strict sulfur limits significantly reduce biodiversity harm. In contrast, MISO's coal-heavy mix raises SO$_2$ emissions above the U.S. average despite only moderate CO$_2$.
Taiwan, dominated by gas, is more carbon-intensive but benefits from tight acid-gas controls, resulting in lower biodiversity impact than MISO.
Relocating the system to QC's nearly all-hydro grid cuts OBI to $\approx1.5\times10^{-4}\,\text{species·yr}$—comparable to EBI and about two orders lower than fossil-heavy grids.

\takeawaybox{ 
Focusing solely on CO$_2$ is not enough—strict SO$_2$/NO$_x$ limits and renewable-heavy grids can cut a supercomputer's total biodiversity impact by an order of magnitude. This highlights the need for pollutant-specific policies.
}



%% file: tex/5-conclusion.tex
\section{Conclusion}

This paper presents the first systematic study of computing’s biodiversity impact across the device lifecycle. By introducing EBI and OBI and developing the \SYSTEM{} framework, we provide a foundation for connecting workloads to biodiversity outcomes. As computing demand grows, biodiversity should become a first-class metric in sustainable system design and optimization.

\begin{acks}
We thank the anonymous reviewers for their valuable feedback.
We acknowledge GCP for providing generous cloud research credits that supported the cloud CPUs used in this study.
This work is supported by NSF CCF-2413870.
\end{acks}